\documentclass[aps,prb,reprint,amsmath,amssymb,superscriptaddress]{revtex4-1}
\usepackage{xcolor}
\usepackage{graphicx}
\usepackage{subfigure}
\usepackage{comment}
\usepackage{hyperref}
\usepackage{braket}
\usepackage[normalem]{ulem}
\usepackage{siunitx} 

\graphicspath{{./figures/}}

\begin{document}

\title{The liquid-liquid phase transition of hydrogen and its critical point: Analysis from ab initio simulation and a machine-learned potential}
\author{Mathieu Istas}
\affiliation{Univ. Grenoble Alpes, CNRS, Institut Néél, 38042 Grenoble, France}
\affiliation{Univ. Grenoble Alpes, CNRS, LPMMC, 38000 Grenoble, France}
\author{Scott Jensen}
\affiliation{Grainger College of Engineering, University of Illinois, Urbana, Illinois 61801, USA}
\author{Yubo Yang}
\affiliation{Center for Computational Quantum Physics, Flatiron Institute, New York, New York 10010, USA}
\affiliation{Department of Physics and Astronomy, Hofstra University, Hempstead, New York 11549, USA}
\author{Markus Holzmann}
\affiliation{Univ. Grenoble Alpes, CNRS, LPMMC, 38000 Grenoble, France}
\author{Carlo Pierleoni}
\affiliation{Department of Physical and Chemical Sciences, University of L'Aquila, Via Vetoio 10, I-67010 L'Aquila, Italy}
\author{David M. Ceperley}
\affiliation{Grainger College of Engineering, University of Illinois, Urbana, Illinois 61801, USA}
\date{\today}
\begin{abstract}
We simulate high-pressure hydrogen in its liquid phase close to molecular dissociation using a machine-learned interatomic potential. The model is trained with density functional theory (DFT) forces and energies, with the Perdew-Burke-Ernzerhof (PBE) exchange-correlation functional. We show that an accurate NequIP model, an E(3)-equivariant neural network potential, accurately reproduces the phase transition present in PBE. Moreover, the computational efficiency of this model allows for substantially longer molecular dynamics trajectories, enabling us to perform a finite-size scaling (FSS) analysis to distinguish between a crossover and a true first-order phase transition. We locate the critical point of this transition, the liquid-liquid phase transition (LLPT), at 1200-1300 K and 155-160 GPa, a temperature lower than most previous estimates and close to the melting transition.
\end{abstract}
\maketitle

\section{Introduction}

Despite being the simplest element of the periodic table, the phase diagram of hydrogen is still only partially known. Understanding hydrogen's behavior across a wide range of temperatures and pressures, up to hundreds of gigapascals (GPa), is crucial for determining the properties of stellar objects such as stars, gas giants, and gas clouds. Further, going back to the work of Wigner and Huntington~\cite{wigner-huntington_35}, 
the metallization transition at high pressure 
has been a long standing goal of experimental investigations \cite{gregoryanz_mre_20}. 
Hydrogen also holds the promise of high-pressure high-temperature superconductivity~\cite{ashcroft_prl_68}. 
 
Fluid hydrogen is found at pressures below approximately 100 GPa in the insulating molecular state.
At higher pressures and temperatures, it is found in the monoatomic conducting state. It is not yet understood whether there is a phase transition or a smooth crossover between these two states. Recently, much attention and debate have been directed toward the possible liquid-liquid phase transition (LLPT) from an insulating molecular fluid of H$_2$ molecules to a conducting monoatomic fluid. If there is a phase transition, 
a first-order LLPT line that ends at a second-order critical point is expected.

Experimentally, there have been several investigations that report a possible LLPT observed with a sharp increase in optical conductivity~\cite{jiang2020spectroscopic, celliers2018insulator} or optical transmittance~\cite{zaghoo2016evidence}. Similarly, hydrogen samples have been heated by laser pulses\cite{zaghoo2016evidence,ohta2015phase}observing a plateau in the heating curve, consistent with a first-order phase transition. 
In other experiments, high pressure  fluid states of hydrogen or deuterium have been reached using shock waves; measurements of electrical resistivity show a rapid but continuous decrease \cite{weir1996metallization} 
above 90 GPa and temperatures higher than 2200K, while density measurements~\cite{fortov2007phase} indicate a phase transition. Interpretations of these experiments are debated~\cite{goncharov2017comment, howie2017comment}.
Experimental measurements at pressures higher than 100 GPa come with many associated difficulties and make a direct LLPT observation challenging. Static diamond anvil cell (DAC) experiments for sufficiently high pressures and  for high enough temperatures are not possible since under these conditions hydrogen will cause the diamonds to break. 
Moreover, the measured quantities are often only indirectly related to the LLPT.
Currently, there is no unequivocal experimental evidence that the LLPT exists\cite{bonitz2024first}.

On the theoretical side, ab initio Molecular Dynamics (AIMD) with forces computed using density functional theory (DFT) have been used to probe the nature and location of the LLPT line. A first-order LLPT has been predicted by DFT investigations~\cite{morales2010evidence, lorenzen2010first, bonitz2024first} and by the Coupled Electron-Ion Monte Carlo\cite{pierleoni2016liquid} based on more accurate quantum Monte Carlo energies, as well as by 
molecular dynamics driven by variational Monte Carlo forces \cite{Mazzola18}. However, determining the precise location of the LLPT line remains a difficult problem.
Within density functional theory, the location and even existence of the LLPT could depend on the choice of functional. 
For example, in DFT functionals such as PBE which have a too small band gap that would favor the atomic liquid with respect to the molecular liquid, 
the LLPT could be suppressed~\cite{vorberger2007hydrogen,holst2008thermophysical}
~\footnote{This suppression of LLPT might also be an effect of using a single k-point rather than averaging over k-points, in other words a finite-size effect.}.

In this work, we restrict the scope of investigation from actual hydrogen to the understanding of hydrogen modeled with a Perdew-Burke-Ernzerhof (PBE) density functional and using classical protons. We will refer to this as {\it PBE-hydrogen}. 
We make this assumption for several reasons. First, simulations using classical hydrogen ignoring the quantum motion of the protons and using the PBE functional are much easier to perform. Second, there are numerous previous works using the PBE functional that we can compare with.  Although different DFT functionals provide better approximations for hydrogen 
\cite{Clay2016}, the LLPT in PBE-hydrogen has been seen previously~\cite{morales2010evidence, karasiev2021liquid, Bryk2020}.
Finally, it has been found that the melting point of PBE-hydrogen is at a lower temperature than actual hydrogen\cite{niu2023stable}. Having a possible stable crystal phase complicates the analysis of the LLPT. 
Even though evidence for the LLPT as been observed in PBE-hydrogen, no systematic methodology has been used to locate the phase transition line and its critical point. 

Here, using recent state-of-the-art machine-learned interatomic potentials (MLIP) to handle large system sizes and simulation times, we show that the LLPT is observed with an E(3)-equivariant neural network potential. We perform a finite-size scaling (FSS) analysis to establish the first-order nature of the phase transition and to locate the critical point for PBE-hydrogen and estimate its relation to the melting transition. 

In the following section we discuss general aspects of establishing the order of a transition with simulation, and in Section \ref{sec:mlip_hydrogen} our machine learning model. Section \ref{sec:existence_llpt} concerns the results of the ML model for the LLPT and melting and compares with previous work. Finally, we reach some conclusions. The Appendices contain more details of our study.

\section{Appearance of the LLPT in a finite-size system}

The nature of the LLPT is difficult to determine in finite-size simulations of PBE-hydrogen. In previous work with this model, the first-order behavior of the LLPT is often identified by a kink in the equation of state, a change in the radial distribution function, and sometimes by less fundamental quantities that rely on heuristics, such as the number of molecules. 
Numerical studies are typically performed with a single finite-sized system, and observe discontinuities in instantaneous properties, e.g. in the density and radial distribution function. However, sufficiently long simulations will display rapidly varying but actually smooth averages, making it difficult to distinguish between a phase transition and a cross-over. Hysteresis by itself does not ensure the first-order character of the phase transition in a finite-size system. For example, both the Ising model and the Lennard-Jones potential exhibit magnetization or density switches between phases in finite systems even at the critical temperature~\cite{binder1984finite, binder1987finite, wilding1997simulation}.

The LLPT is thought to belong to the same universality class of these systems since it has a scalar order parameter, i.e. the density variation between the atomic and molecular phases caused by the different volumes occupied by atoms versus molecules.
In the thermodynamic limit, a discontinuity in the density as a function of pressure defines a first-order transition. At the critical point, the
transition is continuous, but non-analytic, characterized by power laws with known scaling exponents. 

To determine whether or not there is a phase transition, a standard approach tested on many lattice models is to use finite-size scaling~\cite{binder1984finite, wilding1997simulation, newman1999monte}. This can distinguish a weak first-order phase transition from a second-order one \cite{li2024comprehensive}.
Finite size scaling consists in simulating many different sizes at the phase transition. From the results, one can then extrapolate the critical temperature, pressure or exponents to the thermodynamic limit. 

The reason why finite size scaling has not been applied to the LLPT problem is simply because of the significant computation cost of first principle methods, and its scaling, typically $N^3$ for DFT. That renders simulation of large systems prohibitive.
Finite-size scaling has been primarily applied to simple lattice models. The most complex  continuum model previously studied is the Lennard-Jones potential, a pair potential that is too simple to describe hydrogen.
Since numerical simulations of high pressure hydrogen require expensive numerical methods, this constrains the simulations to short time scales with a limited number of atoms.
Careful estimation of phase transition boundaries requires on the order of a decade of system sizes with well-converged simulations.  Critical slowing down can be expected near the phase transition.

We can do simulations of much larger systems with machine learning interatomic potentials (MLIP).
These have been introduced more than 15 years ago \cite{behler2007generalized}, and have improved greatly since then \cite{batzner2023nequip, batatia2022mace} with breakthroughs in recent years. 
These potentials are empirically trained using energy, forces and stresses obtained from ab-initio data.
 Despite having only a few physical principles encoded into them, they are able to reproduce the potential energy surface (PES) and thereby the dynamics of the computationally expensive ab-initio simulations.
Having a large number of parameters, these models are more flexible compared to the more traditional effective potentials, giving a good trade-off between speed and accuracy; faster than ab-initio simulations, but still able to simulate complex realistic materials and quantitatively reproduce their various physical quantities.

\begin{figure}[bth]
\includegraphics[scale=0.5]{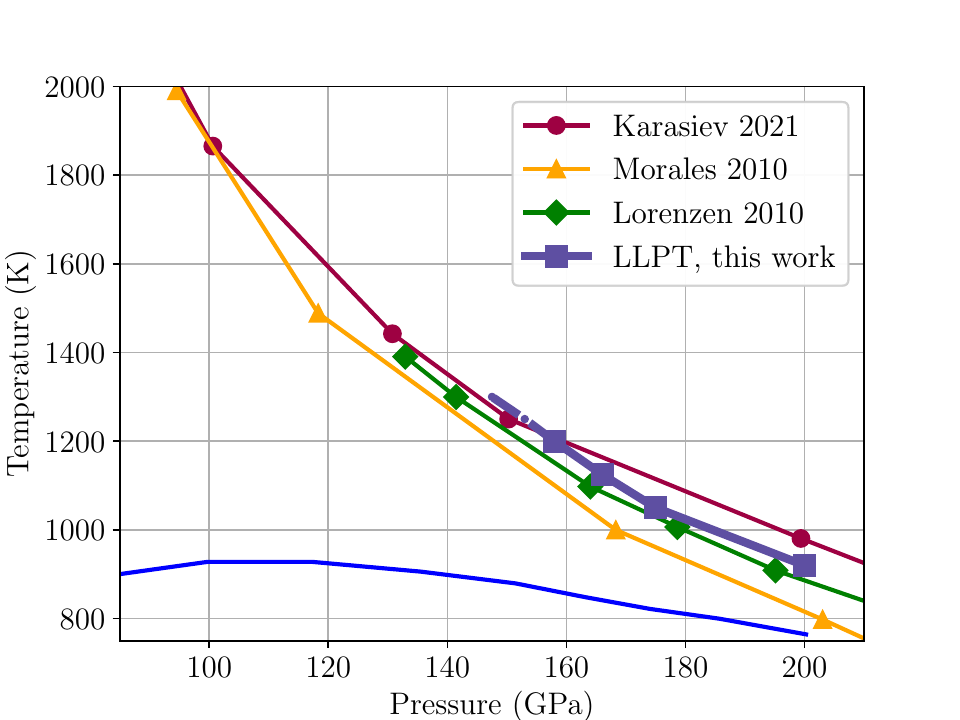}
\caption{Phase diagram of PBE-hydrogen obtained from different calculations.
The lines with symbols correspond to estimations of the LLPT.  
The dark blue curve with squares corresponds to our estimation, the dashed part of the line represents our estimation of the uncertainty in the critical point.  Karasiev  et al.\cite{karasiev2021liquid} did not give an estimation for the critical point, and Morales et al.\cite{morales2010evidence} gave an estimation of about 2000 K.
The lower blue line is an estimate of the PBE-hydrogen melting transition\cite{morales2010evidence}. 
}
\label{fig:phase_diagram_llpt}
\end{figure}

However, in practice, the use of MLIPs require caution due to the ``black-box" nature of this approach. There is a risk that models that are not carefully tested will add more uncertainty to an already controversial subject \cite{Cheng2020,karasiev2021liquid}.
In this work, we use a model based on state-of-the-art message passing neural networks, Nequip \cite{batzner2023nequip}  and test it extensively to prove that this model can be a trustworthy tool to simulate dense hydrogen with first-principle accuracy.
We go beyond the usual comparison of forces and energies errors \cite{batzner2023nequip, batatia2022mace} and make direct comparison between ab-initio and MLIP driven molecular dynamics, as advised in the benchmark proposal in Ref.~\onlinecite{fu2022forces}.
It has even been recently proposed that the LLPT, as it is noticeably hard to simulate and even state-of-the-art machine learning models fail to reproduce, is an appropriate benchmark to compare models \cite{bischoff2024hydrogenpressurebenchmarkmachinelearning}.
In this study we push it further, by comparing a MLIP that is trained on a small number of particles (96 atoms) to AIMD simulations of systems of between 200 and 2048\cite{karasiev2021liquid} hydrogen atoms.
The LLPT phase diagram we obtain in this work from MLIP compared to previous AIMD results is given in fig.~\ref{fig:phase_diagram_llpt}.

\section{Machine learning interatomic potentials for hydrogen}
\label{sec:mlip_hydrogen}

To train the MLIP we use the publicly available\cite{QMC-hamm} database of hydrogen configurations described in Ref.~\onlinecite{niu2023stable}.
The database contains hydrogen configurations with 96 atoms in both the solid and liquid states at pressures between 50 and 200 GPa and at temperatures between 600 and 2000 K.
The current study 
used 54,000  configurations with the energies, forces and stresses computed using the PBE functional.

MLIPs do not make any assumption about the PES, except for symmetry, locality and smoothness.
Locality is enforced by writing the total energy as a sum of $N$ local energies, where $N$ is the number of atoms.
Each proton's local energy is a function of the positions of the neighboring protons within a certain cutoff radius.
This cutoff radius, an important parameter of the model, was fixed to $ r= 2.5$ \AA{} in this study, a choice motivated in App.~\ref{sec:MLIP_parameters}.

In this work we use Nequip\cite{batzner2023nequip}, an architecture that improved MLIP models with the treatment of physical symmetries using equivariant graph neural networks.
The Nequip model was trained for 100 epochs, 48,000 training configurations randomly selected, 2,000 samples kept for validation and the remaining used for testing.
The mean absolute error averaged on the 4,000 testing samples is \SI{1.94}{\milli\eV} per atom on energy, \SI{170}{\milli\eV\per\angstrom} on forces and \SI{525}{\milli\eV\per\cubic\angstrom} on stresses.

\label{sec:model_validation}

Training is performed by minimizing the weighted sum of errors on energies, forces and stresses.
While low energy and forces errors are preferable, low errors alone are not enough to reproduce correct thermodynamic quantities \cite{fu2022forces}, and even small force errors on the order of  \SI{5}{\milli\eV\per\angstrom} can lead to unstable dynamics.
These errors are only a proxy for the real quantities of interest.
What matters in a MLIP is its ability to reproduce the Equation of State (EOS), the radial distribution function, and dynamical quantities such as diffusion.

In the following, we use the dimensionless spacing between protons, the Wigner-Seitz radius, $r_s$ defined as $(4 \pi/3 )r_s^3 =v$ where $v$ is the volume per proton in units of bohr$^3$ to specify the density of the system.
This can be the instantaneous density or the overall time averaged density, depending on the context.

\begin{figure}[tbh]
\includegraphics[width=\linewidth]{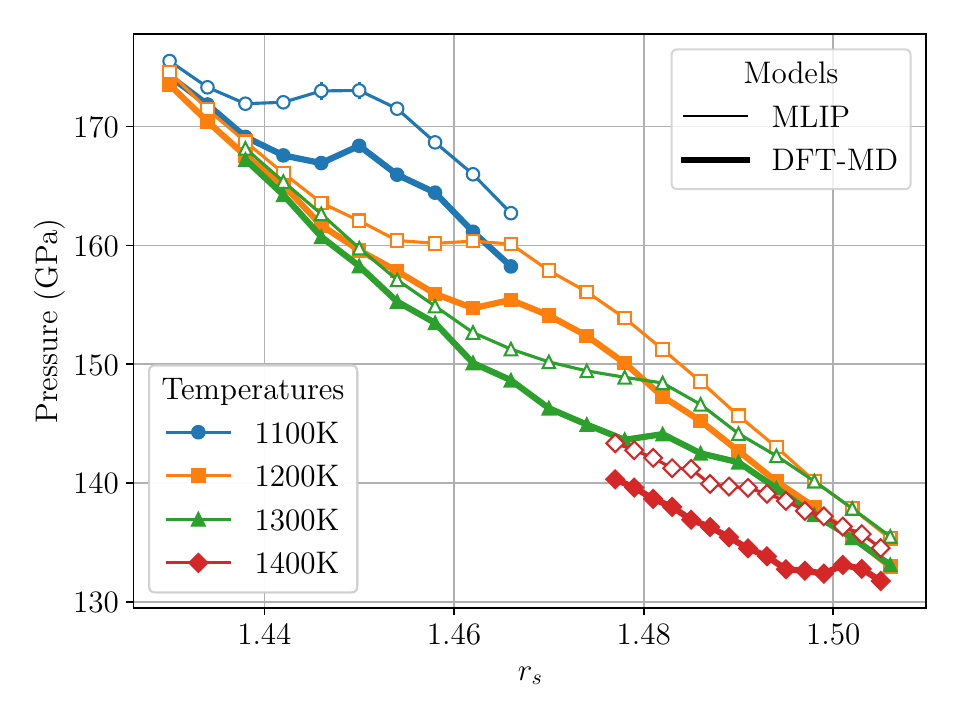}
\caption{EOS for a 200 atom system at 4 temperatures between 1100K and 1500K. Simulations are performed at constant NVT. Open symbols show the MLIP model and filled symbols the DFT simulations. Statistical errors outside the transition region are smaller than the symbol size.}
\label{fig:pressure_PBEMD}
\end{figure}
\begin{figure}[tbh]
\includegraphics[width=\linewidth]{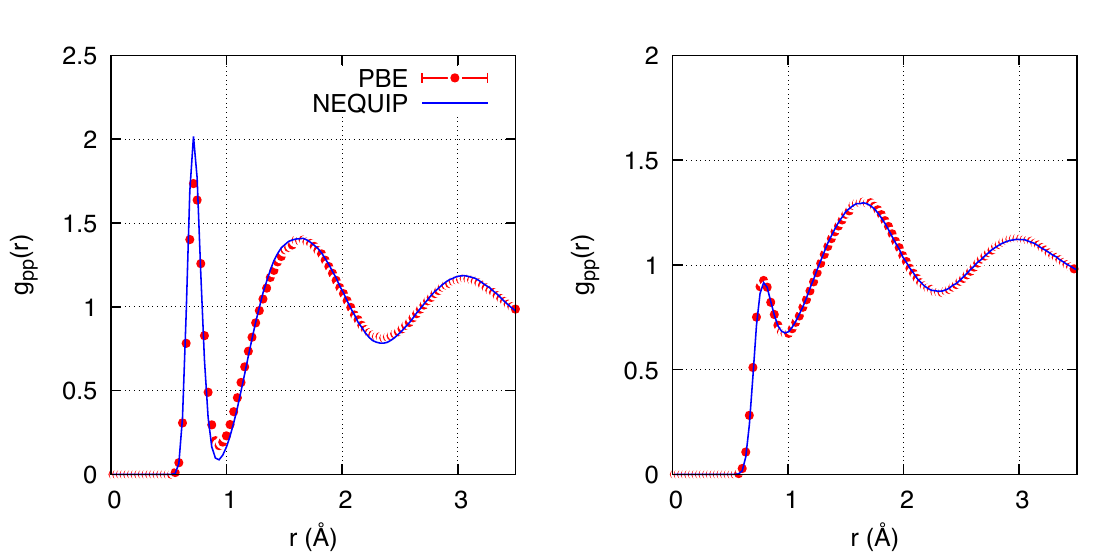}
\caption{
Comparison between proton-proton radial distribution functions from PBE (closed circles) and Nequip (lines) models for the systems of 200 atoms at T=1100K. Left panel: $r_s = 1.466$ (0.855 $\rm{g} \rm{cm}^{-3}$) in the molecular phase,right panel: $r_s = 1.43$ (0.922 $\rm{g} \rm{cm}^{-3}$) in the atomic phase.
PBE trajectories were 100fs and Nequip trajectories 250ps durations respectively.
}
\label{fig:PBE_MLIP_comparison_gr}
\end{figure}

For hydrogen, a machine learning potential with an error of \SI{5}{\milli\eV\per atom} and \SI{300}{\milli\eV\per\angstrom} failed to reproduce the LLPT present in the dataset \cite{karasiev2021liquid}.
To validate our model, trained on small systems, we compare its performance against ab-initio molecular dynamics simulations on larger systems.
The comparison of the EOS for 200 atoms is shown in fig.~\ref{fig:pressure_PBEMD}.
The EOS matches in the atomic phase (small $r_s$). On the molecular side (larger $r_s$), the MLIP model has a larger pressure by about 4GPa. Some of the differences in the transition region could be due to inadequate convergence of the AIMD simulations.

Quantitative agreement is also observed on local correlations, as shown in fig.~\ref{fig:PBE_MLIP_comparison_gr} for the radial distribution functions between protons.
As for the EOS, we observe a very good agreement between PBE and Nequip models in the atomic side of the transitions (right panel) while in the molecular side (left panel) the agreement is less good. The difference in heights of the first maximum and the first minimum is of particular concern. 

An additional benchmark is mandated since finite-size scaling theory relies on the simulation of different sized systems.
In App.~\ref{sec:DFT_MD_dynamics}, we further compare AIMD with systems of $N=200$ atoms and $N=2048$ atoms from the study descrived in ref.~\onlinecite{karasiev2021liquid}.
Despite the imperfections, the previous simulation-based benchmarks showed quantitatively good agreement, encouraging us to use the model for our LLPT study.

\section{Existence, location and order of the liquid-liquid phase transition}
\label{sec:existence_llpt}
To determine the location of the LLPT line and the critical point from the EOS,
we run simulations in the isothermal-isobaric NPT ensemble using periodic boundary conditions.
All simulations use the Nequip MLIP,  last for $200$ ps or longer with a timestep of 0.5 fs. 
To prove our model is not exhibiting a crossover, we use finite-size scaling theory and study the EOS for different system sizes. See chapter 8 of ref.~\onlinecite{newman1999monte} for an introduction to this method.

\subsection{First-order line and critical point}

\subsubsection*{Equation of state and compressibility}

We first prove the first-order nature of LLPT by showing that the average density is discontinuous in the thermodynamic limit.

\begin{figure}[tbh]
\includegraphics[width=\linewidth]{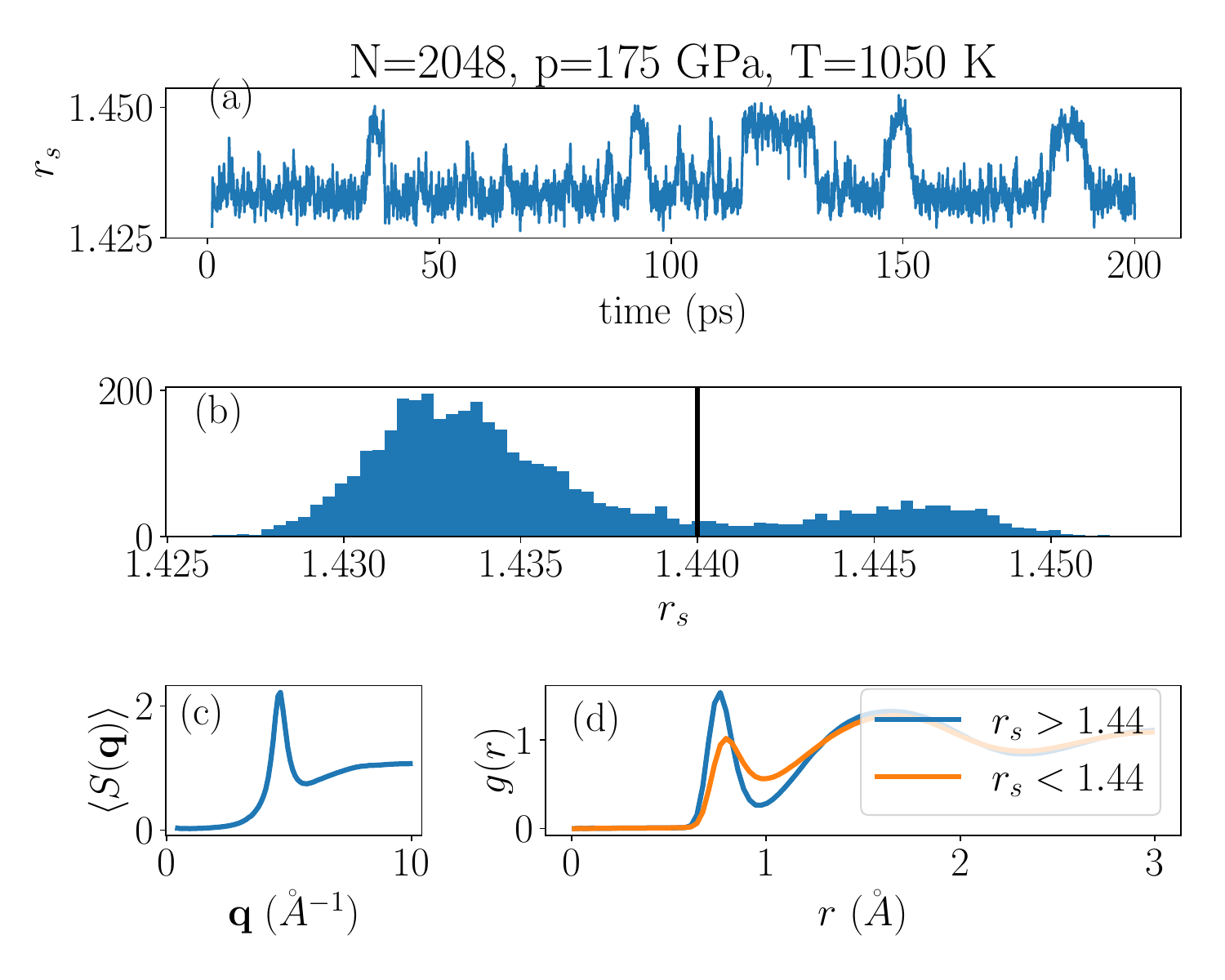}
\caption{(a) Instantaneous $r_s$ value as a function of the simulation time for 2048 particles at $P = 175$ GPa and $T = 1050$ K.
(b) $r_s$ histogram for the same simulation. The vertical line at $r_s = 1.44$ indicates the threshold used in panel (d).
(c) Structure factor averaged over all time steps. (d) Radial distribution function computed for configurations with $r_s>1.44$ (blue line) and configuration with $r_s<1.44$ (orange line). 
}
\label{fig:density_traj_2048_175_1050}
\end{figure}

Similar to the magnetization of the Ising model\cite{binder1984finite}, the density near the phase transition exhibits sharp changes, ``jumping" back and forth between the two different phases, as shown for 2048 atoms in fig.~\ref{fig:density_traj_2048_175_1050}.
The simulation was done under conditions that are consistent with the LLPT seen in  Morales et al.\cite{morales2010evidence} and Karasiev et al.\cite{karasiev2021liquid}. Note that the time between jumps is rather long for this system. Only five jumps occurred in 200 ps, which makes an ab-initio study of the transition in this 2048 atom system costly.
We show a direct comparison with the latter in fig.~\ref{fig:Karasiev_EOS}.
The histogram of the density shown in fig.~\ref{fig:density_traj_2048_175_1050}(b) has two modes. 
The radial distribution function is computed in fig.~\ref{fig:density_traj_2048_175_1050}(d) for each mode separately.
The increase in the height of the first peak in $g(r)$ shows that the high density phase ($r_s<1.44$) is atomic and the low density phase ($r_s>1.44$) is molecular. Figure.~\ref{fig:density_traj_2048_175_1050}(c) also shows the structure factor, proving that the system is in the liquid state.

\begin{figure}[h]
\begin{minipage}{.32\linewidth}
\centering
\includegraphics[width=\linewidth]{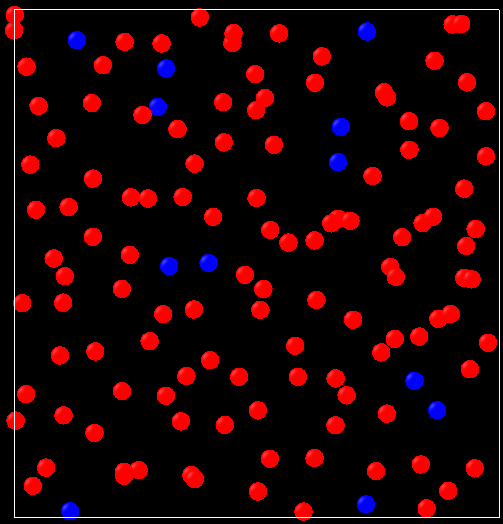}\\
(a) t=$10.2$ ps
\end{minipage}
\begin{minipage}{.32\linewidth}
\centering
\includegraphics[width=\linewidth]{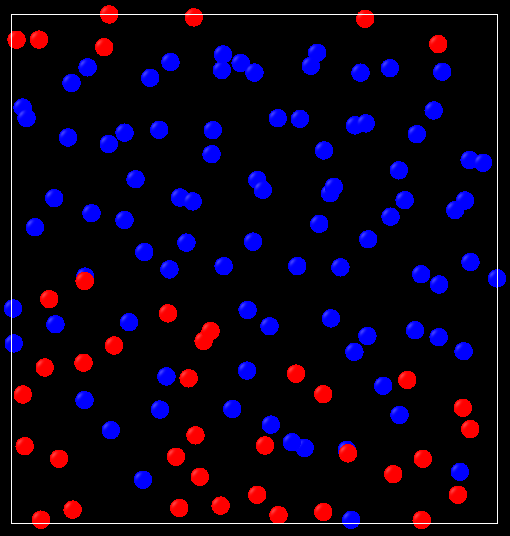}\\
(b) t=$10.5$ ps
\end{minipage}
\begin{minipage}{.32\linewidth}
\centering
\includegraphics[width=\linewidth]{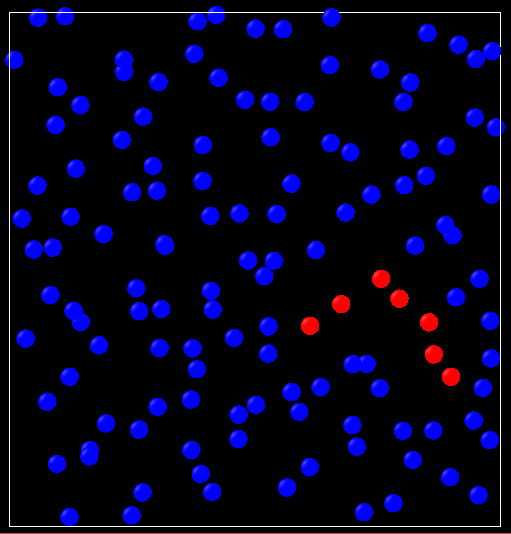}\\
(c) t=$10.9$ ps
\end{minipage}
\includegraphics[width=\linewidth]{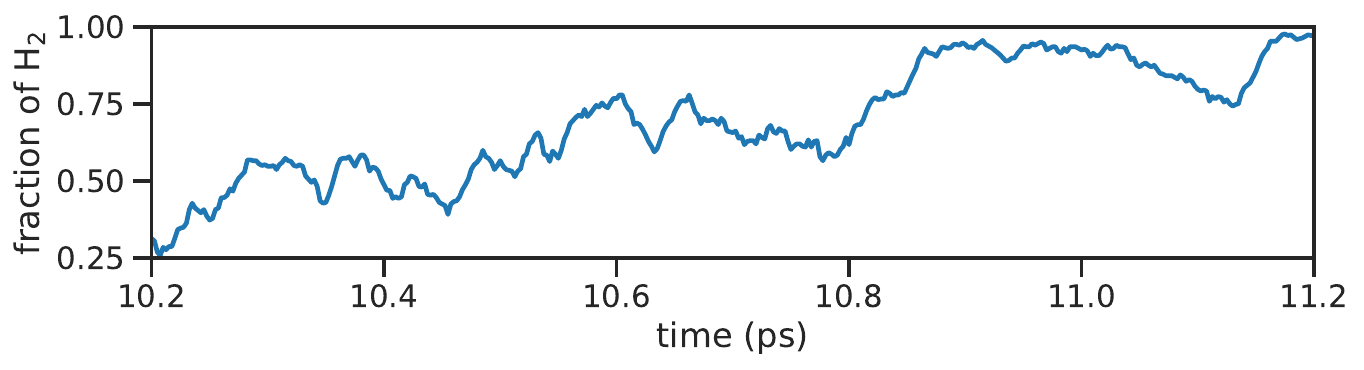}
(d)
\caption{(a)-(c) Three snapshots showing  $2$ \AA~slices of a 768 atom simulation at $P=166$ GPa and $T=1125$ K.   The blue atoms are identified as persistent molecules whereas the red atoms are dissociated.
The number of molecules changes on a timescale of 0.1 ps. (d) Fraction of molecules in the full system as a function of simulation time.}
\label{fig:bond-weight-snapshots}
\end{figure}

In Fig.~\ref{fig:bond-weight-snapshots}, one such jump from the atomic to the molecular phase is visualized.
Slices of three configurations at 1150K from an MD trajectory at the transition pressure are shown.
The atoms that have a persistent molecular bond are colored blue, whereas the unpaired atoms are colored red.
The procedure used to identify molecules is detailed in App.~\ref{sec:persistence}.
During the 0.7 ps shown, the configuration transitions from primarily atomic to primarily molecular. The clustering of the two phases is evident.

\begin{figure*}[tbh]
\includegraphics[width=\linewidth]{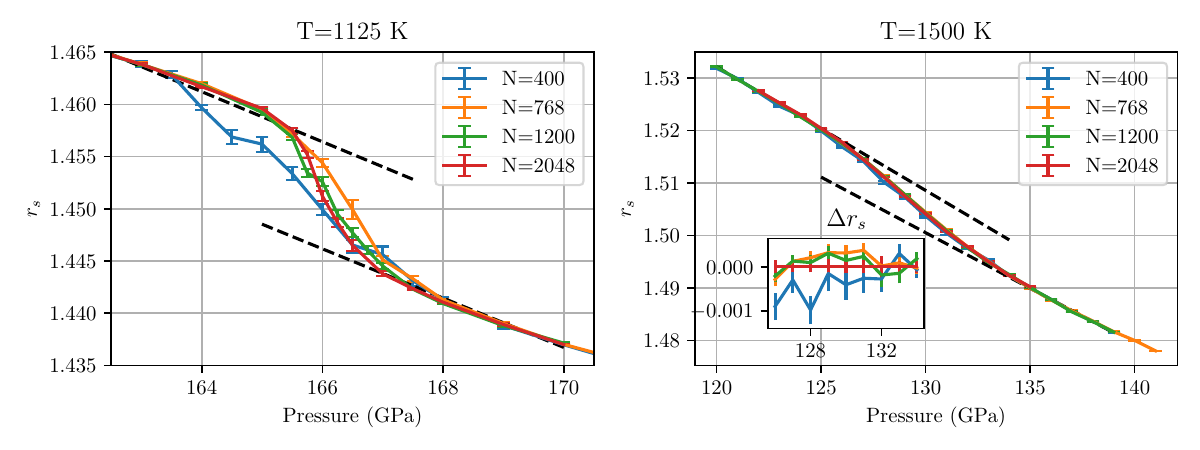}
\caption{ Average density as a function of pressure in $NPT$ simulations for different system sizes using the Nequip MLIP model. Left 1125 K, right 1500 K.  Dashed lines highlight the density difference between and low and high pressure phases. 
In both plots, the 200 atom EOS is not shown for clarity. Statistical error on $\langle \rho \rangle$ is indicated by small bars and do not exceed symbol size on many points.
Inset in the right figure shows the density difference with respect to the 2048 particle density  in the critical region, i.e. $\langle \rho \rangle_N - \langle \rho \rangle_{2048}$.
}
\label{fig:equation_of_state_NPT_1200K_158gpas}
\end{figure*}

As a single system size is insufficient to distinguish a phase transition from a crossover, and although the bimodal histogram of fig.~\ref{fig:density_traj_2048_175_1050} is a strong hint of the presence of a first-order phase transition, we turn to finite-size scaling \cite{newman1999monte} to get more definitive evidence. 

Fig.~\ref{fig:equation_of_state_NPT_1200K_158gpas} illustrates the importance of finite-size scaling to distinguish a phase transition from a cross-over.
We investigate how the average density $\langle \rho \rangle$,
parametrized by $r_s$,
changes with the system size at the phase boundary. The dashed lines are linear fits to the high and low pressure phases shown only to highlight the density differences.
At $T=1125$ K and at pressures near 166 GPa where the density switches between the dashed lines, the slope at its inflection point increases with the system size as expected for a phase transition.
At $T=1500$ K, the density differences between phases is smaller, but still present.
The rapid change in the density as a function of pressure at 1500 K looks similar to the behavior at 1125 K, so that the possibility of a phase transition at T=1500 K could not be dismissed.
But with several sizes, it is seen that the EOS
at T=1500 K does not depend on the system size. PBE-hydrogen undergoes a crossover at this temperature.

To drive the point home, we show that the slope $\frac{\partial \langle \rho \rangle}{\partial P} $ at the transition pressure, $P_t$, not only increases with system size at T = 1125 K, but also diverges in the thermodynamic limit, proving that the density becomes discontinuous.
To do so we fit the following phenomenological equation to the density as a function of pressure at fixed temperature: 
\begin{equation}
\rho(P) = D \tanh(\alpha (P - P_t)) + \kappa (P - P_t) + \rho_t,
\label{eq:tanh_fit}
\end{equation}
This equation is similar to the double Gaussian approximation for the magnetization in the Ising model \cite{binder1981critical, binder1984finite}, where the magnetization is replaced by the density
difference between atomic and molecular liquid and the magnetic field by the pressure difference $P - P_t$. 
The fitting parameters are the density jump between the molecular and atomic phases $D $, a background compressibility $\kappa$, 
the slope $\alpha$ and the offsets $P_t$ and $\rho_t$.

\begin{figure}[tbh]
\includegraphics[width=\linewidth]{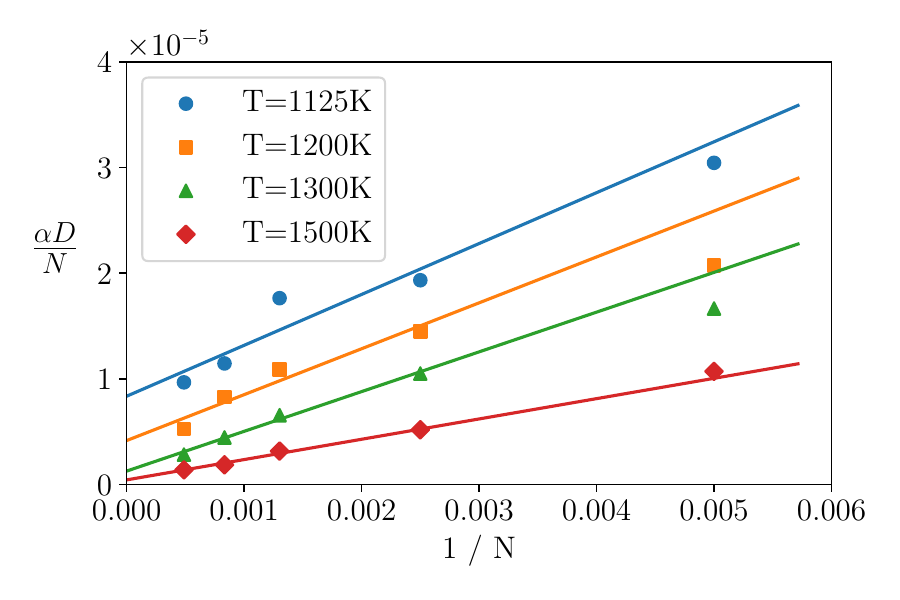}
\caption{ Values of the maximum susceptibility $\alpha D/N $ defined in eqs.~\eqref{eq:tanh_fit} and \eqref{eq:susceptibility_def} as a function of system size.
The lines are least square fits to the four largest system sizes, leaving $N=200$ as an extra point to evaluate the fit quality.
}
\label{fig:slope_N_atoms_T_1125}
\end{figure}

We focus on the first term of eq.~\eqref{eq:tanh_fit} that contains the information about the possible critical behavior.
The susceptibility
\begin{eqnarray}
    \chi_N(P) &\equiv& \frac{\partial }{\partial P} D \tanh(\alpha (P - P_t)),
    \label{eq:susceptibility_def}
\end{eqnarray}
is maximum at $P = P_t$.
In the presence of a first order phase transition, the maximum of $\chi_N(P_t) =  \alpha D$  scales linearly with the number of particles, eventually becoming infinite for an infinite system\cite{binder1984finite}.
Figure~\ref{fig:slope_N_atoms_T_1125} plots $\alpha D  / N$ as a function of $1/N$. 
For system sizes varying from $N=200$ to $N=2048$, the extrapolated value $\lim_{N \to \infty} \alpha D / N$ remains clearly finite for T$=1125$ K,
proving that the system undergoes a first-order phase transition.
It vanishes at $1500$ K.
In between lies the critical point, where the phase transition is second-order
with an {\it a priori} different power law for the divergence of the susceptibility
as a function of $N$.
Our resolution, however, is too low to resolve such changes of the power law. One of difficulties is that
the small density difference  between 
molecular and atomic phases,  $2-3\%$ of the total density at 1125 K, is of 
the same order as the density fluctuations for systems with fewer than a thousand atoms.

\begin{figure*}[tbh]
\includegraphics[width=0.8\textwidth]{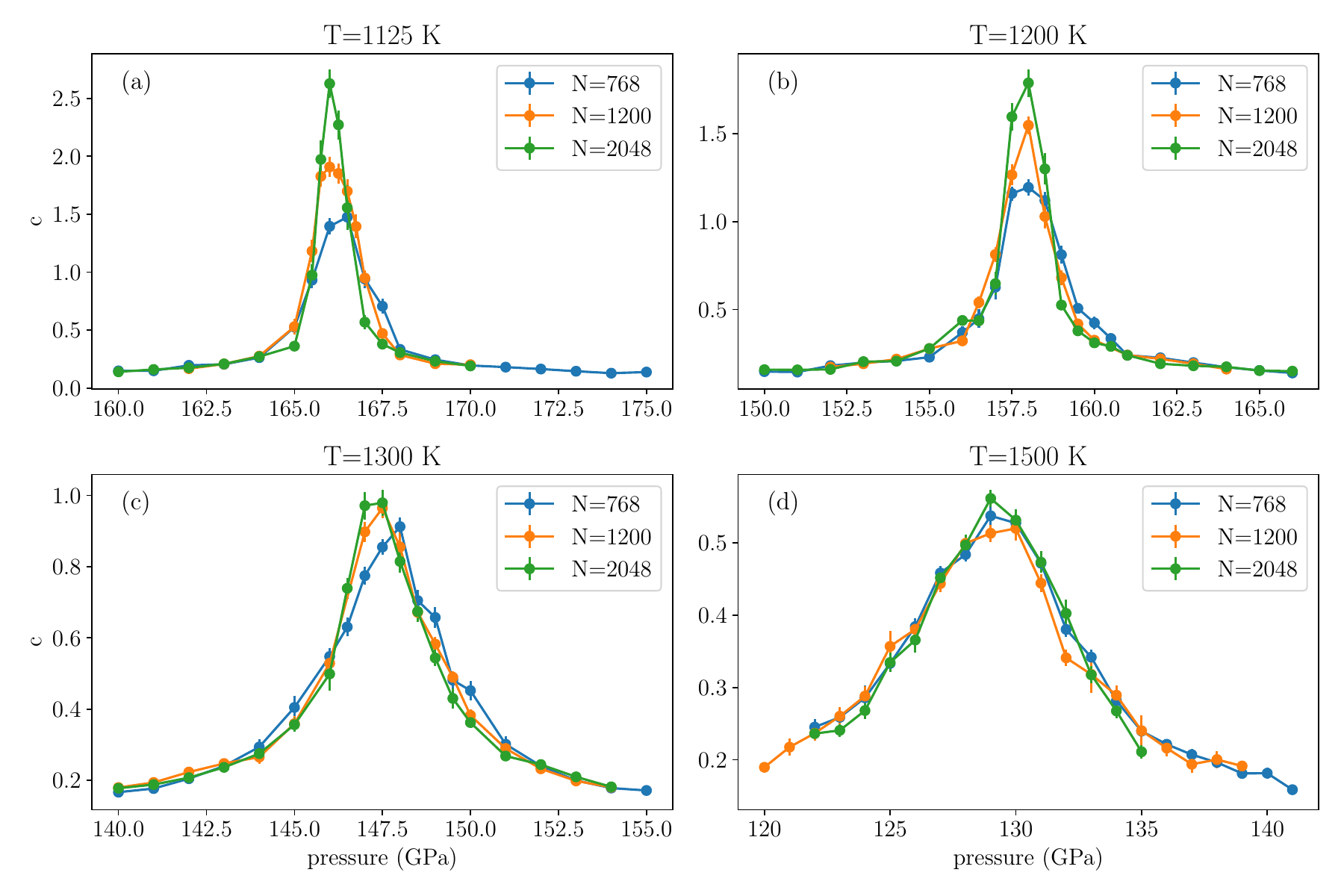}
\caption{Variance of the potential energy per atom, $c$, for $N=768, 1200$ and 2048 atoms at 4 different temperatures from simulations in the $NPT$ ensemble.
}
\label{fig:variance_e_nequip}
\end{figure*}

\subsubsection*{Energy fluctuations}

As an independent check on the critical behavior, we also investigated the fluctuations of the potential energy.
The variance of the total potential energy per unit volume is 
\begin{equation}
    c =(\langle E^2 \rangle - \langle E \rangle ^2)/L^3,
\end{equation}
where $L \sim N^{1/3}$ is the linear extension of the simulation box.
This is the term determining possible non-analytic behavior of the specific heat at a phase transition in the thermodynamic limit.
Away from a transition, the variance of any extensive property is expected to scale with the number of atoms.
A different power law indicates a phase transition, either first-order or continuous.

In fig.~\ref{fig:variance_e_nequip} we plot this variance for several  temperatures and system sizes.
There is a clear increase in $c$ with the system size for both the lowest temperatures $T = 1125$ K and $T = 1200$ K, while the maxima at $T=1300$ K for the largest systems overlap.
But at T=$1500$ K,  $c$ shows no increase  with system size indicating a crossover.
Since at $T = 1200$ K the variance scales linearly with the system size and $c$ shows little or no increase at $T=1300$ K, we estimate the critical point to be between these temperatures, between 1200 K and 1300 K.

\subsubsection*{Finite size scaling at the critical point}
One might expect to obtain a more precise location
of the critical point by the use of  applying known finite-size scaling results from simpler models in the 
same universality class.
Since the order parameter, the change of density from the molecular to the atomic phase,
is a scalar, we expect the phase transition at the
critical point to be described by the same universality class as the 3D Ising model.
However,  the absence of particle-hole symmetry
introduces important corrections to scaling\cite{rehr1973revised}.
A revised form of the finite-size scaling close to critical points in fluid models such as the Lennard-Jones system has been studied extensively by A. D. Bruce and N. B. Wilding \cite{bruce1992scaling}
considering both density and energy density, $e=E/L^3$.
By appropriate scaling with system size,
they managed to match the
singular part of the joint probability distribution of density and energy fluctuations of the Lennard-Jones model to the universal distribution obtained 
from the 3D Ising model\cite{wilding1997simulation, wilding1996finite}.

\begin{figure}[tbh]
\includegraphics[width=\linewidth]{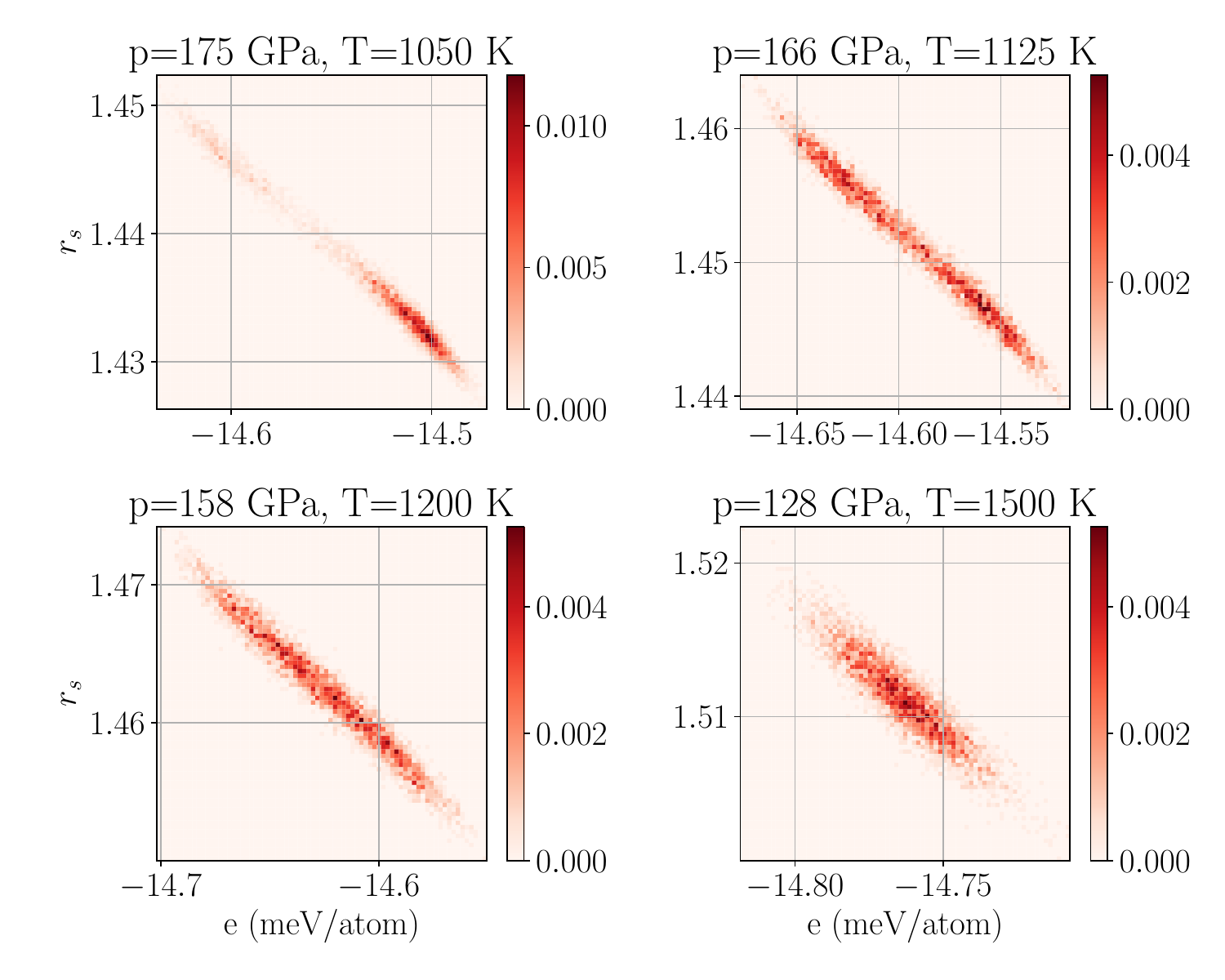}
\caption{Probability distributions $p_N(r_s, e)$ for $N=2048$. Pressures with the maximum variance of the energy at a given temperature are shown. Simulations are at least 250ps long. The energy and density is sampled each 25 fs and is shown as a dot.}
\label{fig:2d_proba_density}
\end{figure}

We plot the joint distribution of density and energy fluctuations $p_N(r_s, e)$ in fig.~\ref{fig:2d_proba_density}. This shows two modes merging as the temperature increases. 
Close to the critical point, 
the non-regular behavior of the distribution which emerges
for large system size, is universal. However, our data
is too sparse to use the finite-size
scaling behavior of the joint distribution.
From Fig.~\ref{fig:2d_proba_density} we may simply expect
the critical point to occur when the behavior of the
distribution function changes between 1125 K and 1500 K.

\subsection{Melting and phase diagram}

\begin{figure}[bth]
\includegraphics[scale=0.54]{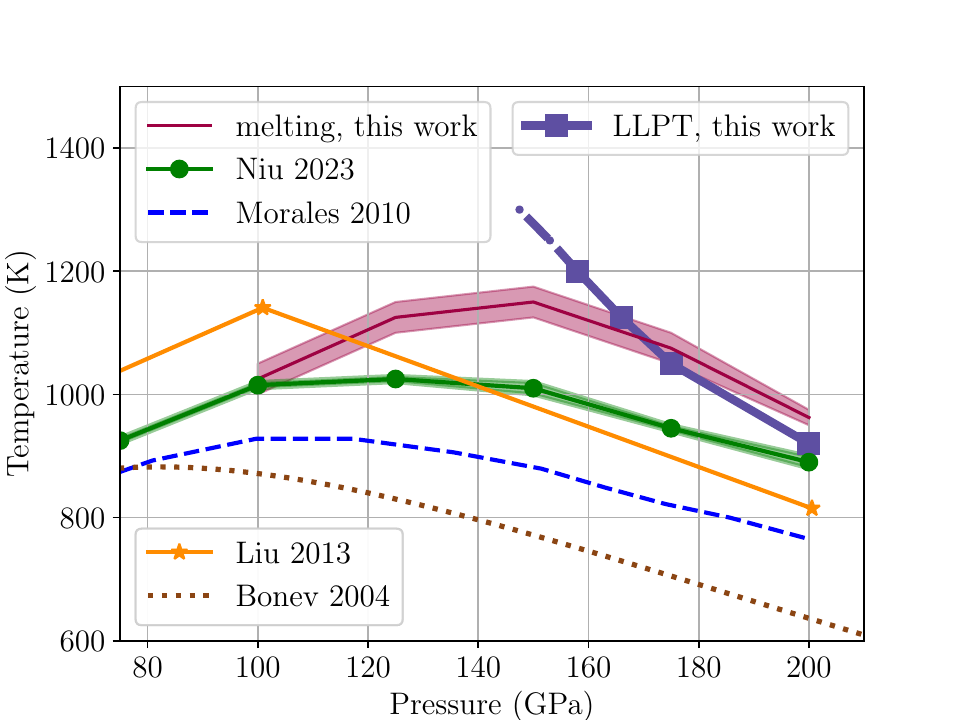}
\caption{Melting and LLPT lines of PBE-hydrogen.
The upper reddish shaded line is our estimation of the melting temperature of the Nequip model, computed using a two-phase method described in appendix ~\ref{sec:Crystallization}.
The purple line with squares is our LLPT estimation from fig.\ref{fig:phase_diagram_llpt}.
The green line with circles is an estimate of the melting transition from a different MLIP (DPMD) trained on the same PBE data\cite{niu2023stable}.
The other lines are estimations of the melting temperature using AIMD with a PBE functional \cite{Liu_anomalous_melting_2013, bonev2004quantum, morales2010evidence}.
}
\label{fig:phase_diagram_melting}
\end{figure}

The melting line obtained with our machine learning model is shown in fig.~\ref{fig:phase_diagram_melting} along with other estimates of the melting transition in PBE-hydrogen.  We note that there is considerable uncertainty of its melting temperature. This uncertainty arises from differences in methods for computing the forces, in the methods used to compute the melting temperature and in the number of atoms used. We note that the later appears quite important. Further studies of the stable crystal structures in this range of temperature and pressures are needed since ref. \onlinecite{niu2023stable} found a transition to an alternate Fmmm-4 phase in a ML model trained on QMC energies. 

All calculations find a maximum for pressures between 80 GPa and 150GPa but with the maximum temperature varying between 820K to 1150K. We note that recent MLIP trained on QMC data\cite{niu2023stable} (i.e. a different model) found a melting temperature above 1700K.  
Having a high melting temperature complicates the understanding of the LLPT. It is possible the solid molecular phase is more stable than either liquid phases so that the LLPT critical point will be inside the solid molecular hydrogen phase.
As shown in App. \ref{sec:DFT_MD_dynamics} the Nequip model has more molecular ordering than the PBE model which could affect the stability of the solid. Further studies are needed.

The melting line of  the Nequip MLIP was obtained by simulating a system initialized with two phases, one half being in the molecular h.c.p. phase and the other half in a molecular liquid.
Details of this procedure and snapshots of these simulations are shown in  App. \ref{sec:Crystallization}.
As shown in fig.~\ref{fig:phase_diagram_melting}, we find that the LLPT for temperatures less than $T=1100$ K is likely below the melting line: the liquid states would be metastable.  
We monitored the structure factor of each simulations, e.g. see fig.~\ref{fig:density_traj_2048_175_1050} to ensure that all runs remained liquid.
Despite the (P,T) conditions of our study allowing for a stable crystal, we found only liquids in our simulations of the LLPT.
We think  the runs performed in this study were not long enough for the system to freeze and that the boundary conditions disfavored crystals. 
The range of temperatures where the LLPT exists and the liquid state are stable seems to be in the narrow range $1100 \lesssim T \lesssim  1250$ K.

It is quite likely that the LLPT meets the Nequip melting line at a triple point around 170 GPa. The melting line for pressures greater than this is between the molecular crystal and the atomic fluid. However, the enthalpy of the three phases: molecular solid, molecular liquid and atomic liquid, are very close.  Further study is needed to make a definitive phase diagram of PBE-hydrogen. As is the case with the LLPT in water \cite{gartner2022liquid, dhabal2024liquid}, we can still study the LLPT as metastable liquid phases. 

\subsection{Comparison with previous works}
The LLPT has been predicted in many previous works. In Fig. \ref{fig:phase_diagram_llpt}  we show those  for classical hydrogen that use a PBE-DFT potential.
We note that the location of the transition, whether it is first-order or continuous, is in remarkable agreement between the various calculations. 
However, our  prediction for the PBE critical point at $T=1250K \pm 50$ is substantially lower than other estimates\cite{scandolo2003liquid, morales2010evidence, lorenzen2010first}, all of which lie between $1500K$ and $2000K$.
The early work of Scandolo\cite{scandolo2003liquid} cautiously states that a crossover cannot be excluded.

Several effects could explain this discrepancy: (i) different details within DFT such as k-points or cutoffs used, (ii) machine learning not reproducing the DFT energies perfectly, or (iii) finite-size effects due to a small number of particles. We believe our conclusions differ from earlier works mostly in the use of finite-size scaling to identify the transition order. 
Had we relied simply on the EOS of a single size system, we would have obtained similar conclusions to earlier works, even  if we had used other properties such as the number of molecules as an order parameter.
In App.~\ref{sec:DFT_MD_dynamics} we show that a machine learning model trained with 96 particles is able to reproduce AIMD simulations with 200 and 2048 atoms, ruling out (i) and (ii) as main sources of errors.
Because of the weak first-order transition in hydrogen, one needs more than a thousand atoms to asses the first-order character of the transition.

\section{Conclusion}
We have presented a finite-size scaling study of the liquid-liquid phase transition of a model of hydrogen, using a machine learning interatomic potential trained on PBE-DFT data.
We found a LLPT in the machine learning model. Finite-size scaling analysis, not relying on details of the order parameter, shows that the transition is consistent with a first-order transition from a molecular fluid to an atomic fluid with a small  density difference between phases.
The location of the transition line in the pressure and temperature phase diagram is consistent with the data used for training and other studies.    
As our results allowed us to differentiate unequivocally a phase transition from a crossover, we found a critical point temperature 1250K $\pm$ 50K substantially lower than previously estimated, bringing it closer to the melting line. 

With well-justified methods tested on a model of hydrogen, future work will apply them to a realistic model\cite{niu2023stable,goswami2024high} of hydrogen and deuterium. Sorting out the relative enthalpies of  the molecular and atomic liquid phases, and the various solid phases is in principle possible leading to definitive predictions of the hydrogen and deuterium phase diagrams. 
\section{Data availability}
The dataset can be downloaded from qmc-hamm website \url{https://qmc-hamm.hub.yt/data.html}. The models, machine learning and ab-initio MD trajectories are available on demand. A link to download the model will be made upon publication.

\acknowledgments
We thank Valentin V. Karasiev for sharing DFT simulations with us.
M. I. and M.O. thank V. Olevano for insightful discusions.
Work by D.M.C. and S.J. was supported by the U.S. Department of Energy (DOE), Office of Science, Basic
Energy Sciences (BES) under Award \#DE-SC0020177. 
CP was supported by the European Union - NextGenerationEU under the Italian Ministry of University and Research (MUR) projects PRIN2022-2022NRBLPT CUP E53D23001790006 and PRIN2022-P2022MC742PNRR, CUP E53D23018440001.
Computations presented in this paper were performed using the GRICAD infrastructure (https://gricad.univ-grenoble-alpes.fr), which is supported by the Grenoble research communities and also performed on
the Illinois Campus Cluster, supported by the National Science Foundation (Awards No. OCI-0725070 and No. ACI-1238993), the State of Illinois, the University of Illinois at Urbana-Champaign, and its National Center for Supercomputing Applications.
The Flatiron Institute is a division of the Simons Foundation.

\bibliography{bibliography}

\begin{appendix}    

\section{DFT, machine learning and molecular dynamics parameters}
\label{sec:MD_with_PBE_200_particles}
We perform AIMD simulations with VASP~\cite{hafner2008ab} driven with DFT forces using the Perdew-Burke-Ernzerhof (PBE) exchange correlation functional\cite{perdew1996generalized} and a standard projector-augmented-wave (PAW) pseudopotential\cite{blochl1994projector}. We performed extensive simulations with N=200 protons in both the NVT ensemble and the NPT ensemble for 3-8 ps. Our simulations scan temperatures from 1000K to 1600K and densities of $r_s=1.43$ to $r_s=1.52$. All DFT simulations use a MD timestep of 1 fs, 500 eV energy cutoff, and a $4^3$ k-point grid centered at the $\Gamma$ point. Convergence with k-points was reached in trial parameter simulations. We have performed further shorter simulations testing cutoff parameters up to 700 atom systems. 

MD performed with the machine learning model used a timestep of $0.5$ fs.
The thermostat and barostat damping values were initialized for 100 timesteps.
The standard atomic weight value of 1.008 was used.

\label{app:MD_comparison}

\section{Simulating LLPT with machine learning}
\label{sec:MLIP_parameters}

The MLIPs developed in recent years are flexible and can reproduce an accuracy comparable to AIMD.
On the other hand, dense hydrogen and in particular its LLPT has been acknowledged\cite{bischoff2024hydrogenpressurebenchmarkmachinelearning} as a system hard to reproduce with MLIP.
We made an extensive study of training procedures to find models that would exhibit the LLPT, both qualitatively and quantitatively.
In this section we provide some guidelines for training MLIPs to reproduce the LLPT that may be applicable to other systems.

We could only reproduce the LLPT with equivariant message passing neural networks, namely MACE\cite{batatia2022mace} and Nequip\cite{batzner2023nequip}.
Of the numerous hyperparameters of these models, we found two that seem very relevant to the MLIP quality, the cutoff radius and the ratio of energy to force weight in the loss function.

The interaction cutoff radius $r_c$, restricts interactions in the model to hydrogen atoms closer than this distance.
On one hand, smaller $r_c$ results in faster simulations.
On the other hand, a bigger $r_c$ allows for a bigger receptive field and more complex interactions and should result in more precise simulations.
We found that the choice of $r_c$ is subtle: increasing $r_c$ does lower errors on configurations with the same number of atoms as in the training set, but the same error evaluated on a larger system does not systematically decrease with $r_c$.

\begin{figure}[tbh]
\includegraphics[width=\linewidth]{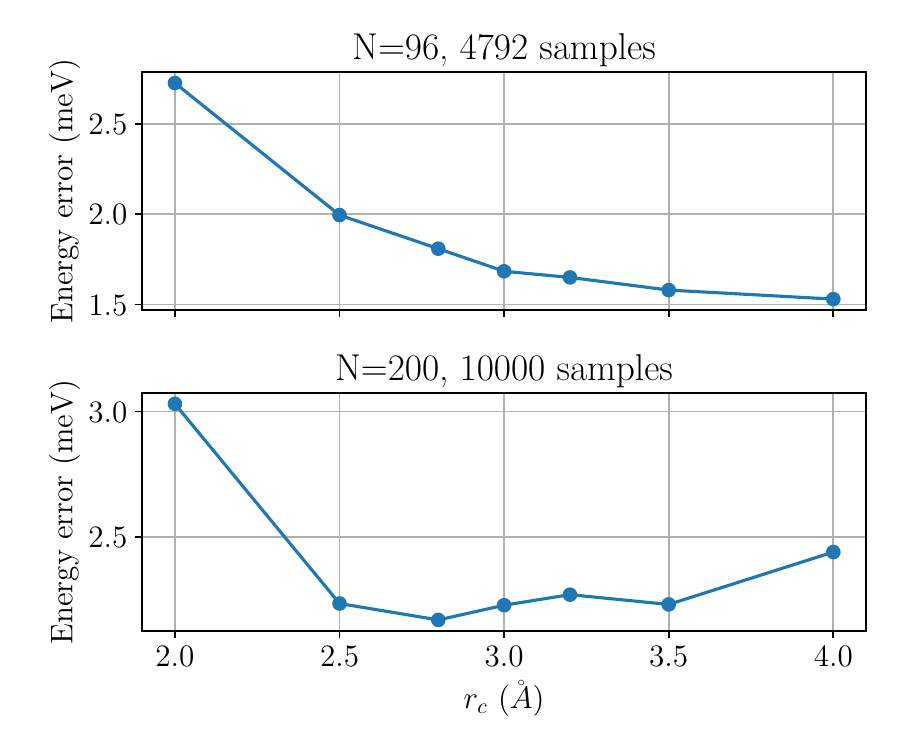}
\caption{ Energy error (MAE) per atom as a function of the cutoff radius: top panel 96 atoms, bottom panel 200 atoms. Each point corresponds to a different Nequip model trained with the same architecture but with a different value of $r_c$. The model with the minimum error in the first 100 epochs in the  96 atom training  was used for both the top and bottom panels.  
}
\label{fig:energy_cutoff_radius}
\end{figure}

This is illustrated in fig.~\ref{fig:energy_cutoff_radius}, which shows the mean absolute error (MAE) on energy per atom between the Nequip MLIP model and the DFT configurations.
The top part shows the average error on various configurations from the test set, each containing 96 atoms
while the bottom part shows the same quantity on configurations of 200 atoms selected from the ab-initio trajectory described in App.~\ref{sec:MD_with_PBE_200_particles}.
Quantitative comparison of the two curves is not possible, as the test set covers a wide range of pressure and temperatures, including solid configurations, while the 200 atom trajectory contains only liquid configurations.
However, it is seen that while increasing $r_c$ systematically reduces the error on the test set the error on the 200 atom trajectory seems to plateau with a slight increase for $r_c > 2.5 $ \AA.
The typical size of a configuration with 96 atoms has an extent of $5-6$ \AA \hspace{0.1mm}, and the receptive field of the machine learning model is bigger than $2r_c$ due to the  message passing mechanism.
A possible explanation for the error increase for 200 atoms is that a value of $r_c$ that is too large might bias the model into learning finite-size effects.

The second important hyperparameters are the energy and force weights ($\lambda_E, \lambda_F$) used in the loss function:
\begin{equation}
\begin{split}
L = \lambda_e \frac{1}{N} | \hat E - E | + \lambda_F \sum_{i=1}^N \sum_{\alpha}^3 \frac{1}{3N} \left| \frac{\partial \hat E}{\partial r_{i, \alpha} } - F_i \right| \\ + \lambda_S \sum_{\mu, \nu} \left | \hat \sigma_{\mu, \nu} - \sigma_{\mu, \nu} \right |.
\end{split}
\label{eq:loss_term}
\end{equation}
The stress tensor is defined as
\begin{equation}
    \sigma_{\mu, \nu} = \frac{1}{V} \frac{\partial E}{\partial \epsilon_{\mu, \nu}},
\end{equation}
where $V$ is the box volume and $\epsilon_{\mu, \nu}$ is the mechanical strain, which corresponds to an elastic deformation of all atomic positions.
We chose the values $\lambda_E = 100$~eV,  $\lambda_F=100~eV/\AA$ and $\lambda_S = 1~eV/\AA^3 $.
Increasing the ratio $\lambda_E / \lambda_F$ during the training will result in low energy errors, but larger force errors.
Empirically, we find that increasing $\lambda_E$ reduces the energy error, while having a small effect on forces error\cite{goswami2024high}.
Even though forces are the quantity driving the MD simulation, thermodynamic quantities such as the free energy rely only on the energy of the model. See the discussion in ref. \onlinecite{goswami2024high}. Hence we increased the value of $\lambda_E$ so that it would dominate the loss function, even if the force errors are slightly higher.
Finally, Nequip allows for a stress error term in the loss function of eq.~\eqref{eq:loss_term}.
We did not systematically investigate $\lambda_S$, but found that having a non-zero stress term helps training and results in better models.

\section{Comparison between DFT and ML molecular dynamics}
\label{sec:DFT_MD_dynamics}

As noted in ref.~\onlinecite{fu2022forces}, low energy and forces errors alone do not imply quantities derived from machine learning molecular dynamics will be close to quantities derived from AIMD.
The current study is even more challenging, as we simulate different system sizes in two different phases. We need to ensure our MLIP trained on configurations with 96 atoms can simulate larger systems with minimal bias containing thousands of hydrogen atoms.

We ran the Nequip model at the same conditions as the AIMD simulations and compare the EOS  as shown in Fig.~\ref{fig:pressure_PBEMD}  for N=200 particles with temperatures from T=1100 K to 1400 K.
We find qualitative agreement with a shift in pressure of about 4 GPa in the molecular phase.
Similarly in figure \ref{fig:PBE_MLIP_comparison_gr} we find a very good agreement with the models for the $g_{pp}(r)$, better in the atomic than in the molecular phase. 

\begin{figure}[tbh]
\includegraphics[width=\linewidth]{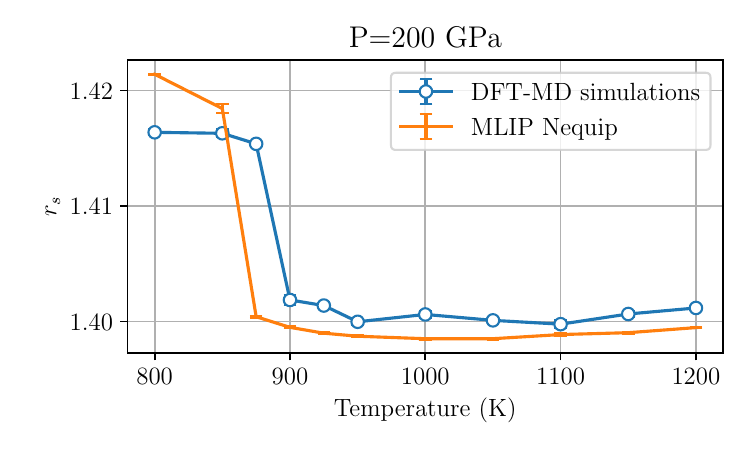}
\caption{Comparison of density versus temperature between AIMD from ref.~\onlinecite{karasiev2021liquid} (blue curve and points) and the Nequip model (orange).
Both simulations used 2048 atoms with an NPT ensemble at 200 GPa.
}
\label{fig:Karasiev_EOS}
\end{figure}

We further test the scaling of our model by comparing it to a 2048 atom ab-initio simulations that V. Karasiev kindly shared with us.
We compare their densities  with our model at 200 GPa versus temperatures from 800 K to 1200 K in fig.~\ref{fig:Karasiev_EOS}.
Both behaviors exhibit a sharp change in density at similar values of temperature.
The statistical error is smaller than symbol size, but is not well estimated in the critical region due to critical slowing down near the phase transition.
The differences between the two curves are likely due to imperfections in the machine learning model, though part of the discrepancy may also arise due the relatively short times in AIMD and to the different 
details within DFT such as k-point grid and the MD parameters such as the timestep, barostat and the initial configurations.

\begin{figure}[tbh]
\includegraphics[width=\linewidth]{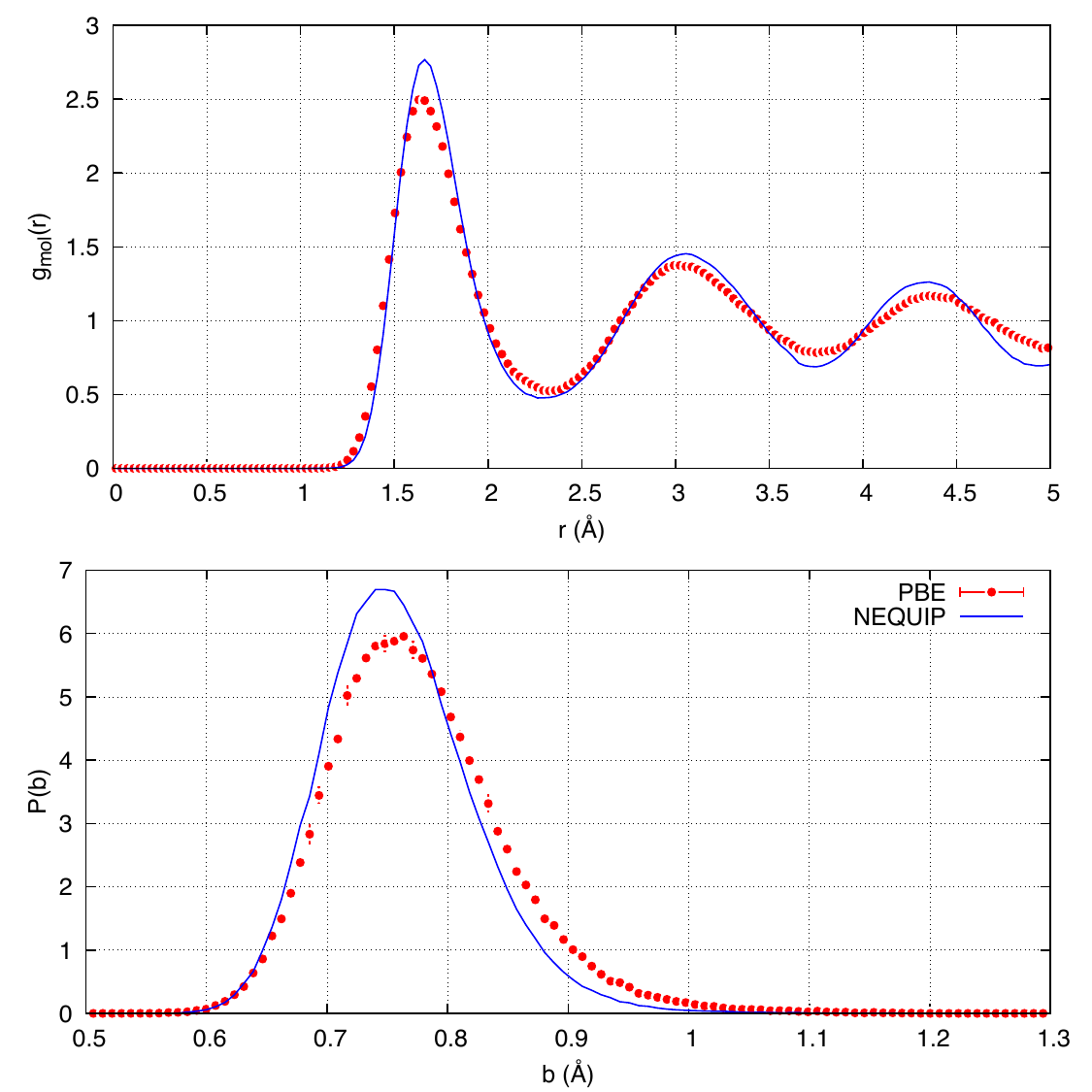}
\caption{Comparison of the molecular radial distribution function (upper panel) and molecular bond length distribution (lower panel) between the AIMD and Nequip models  at $T=1100K,~ r_s=1.466,~ N=200$. 
}
\label{fig:PBE_MLIP_comparison_molecular}
\end{figure}

Finally we investigate the comparison between the models in the molecular phase. For each configuration along the trajectory we detect molecules using a linear assignment problem solver. In figure \ref{fig:PBE_MLIP_comparison_molecular} we compare the molecular pair distribution function and the bond length distribution between the two models. Again we see a good but not perfect agreement between AIMD and Nequip. In particular, Nequip has a stronger molecular character with slightly shorter bond length. 

An important orientational correlation between pairs of molecules is the geometric contribution to the quadrupole-quadrupole interaction. For molecules with inversion symmetry and thus lacking a permanent dipole, the quadrupole term is the first non-zero term in the multipole expansion of their electrostatic interaction.
Considering a pair of molecules with bond unit vectors ${\bf \hat{b}}_1$ and $\hat{\bf b}_2$ whose centers of mass are separated by the vector $\bf{r}_{12}={\bf R}_{1}-{\bf R}_{2}$ with unit vector $\hat{\bf r}_{12}=\bf{r}_{12}/|\bf{r}_{12}|$, the geometric contribution is
\cite{HCB}
\begin{eqnarray}
   \Gamma &=& \frac{1}{8} \left\{1 
    - 5 \cos^2\theta_1 - 5 \cos^2\theta_2 - 15 \cos^2\theta_1\cos^2\theta_2 \right. \nonumber \\
    &+& \left. 2(5 \cos\theta_1 \cos\theta_2 - \cos\phi)^2 
    \right\}, 
    \label{eq:gamma}
\end{eqnarray}
where $\cos \theta_i=\hat{\bf b}_i\cdot \hat{\bf r}_{12}~(i=1,2)$, and $\cos\phi=\hat{\bf b}_1\cdot \hat{\bf b}_{2}$.
The average quadrupole-quadrupole interaction for two molecules at distance $r$ in atomic units is
\begin{equation}
    \mathcal{V}(r)=\frac{6\,\Theta^2}{r^5}\times\Gamma(r).
    \label{eq:Vqq}
\end{equation}
where $\Theta$ is the quadrupole moment of a hydrogen molecule and $\Gamma(r)$ is the average  of the geometric factor of eq. (\ref{eq:gamma}) over molecular pairs  separated by a distance $r$.

\begin{figure}[tbh]
\includegraphics[width=\linewidth]{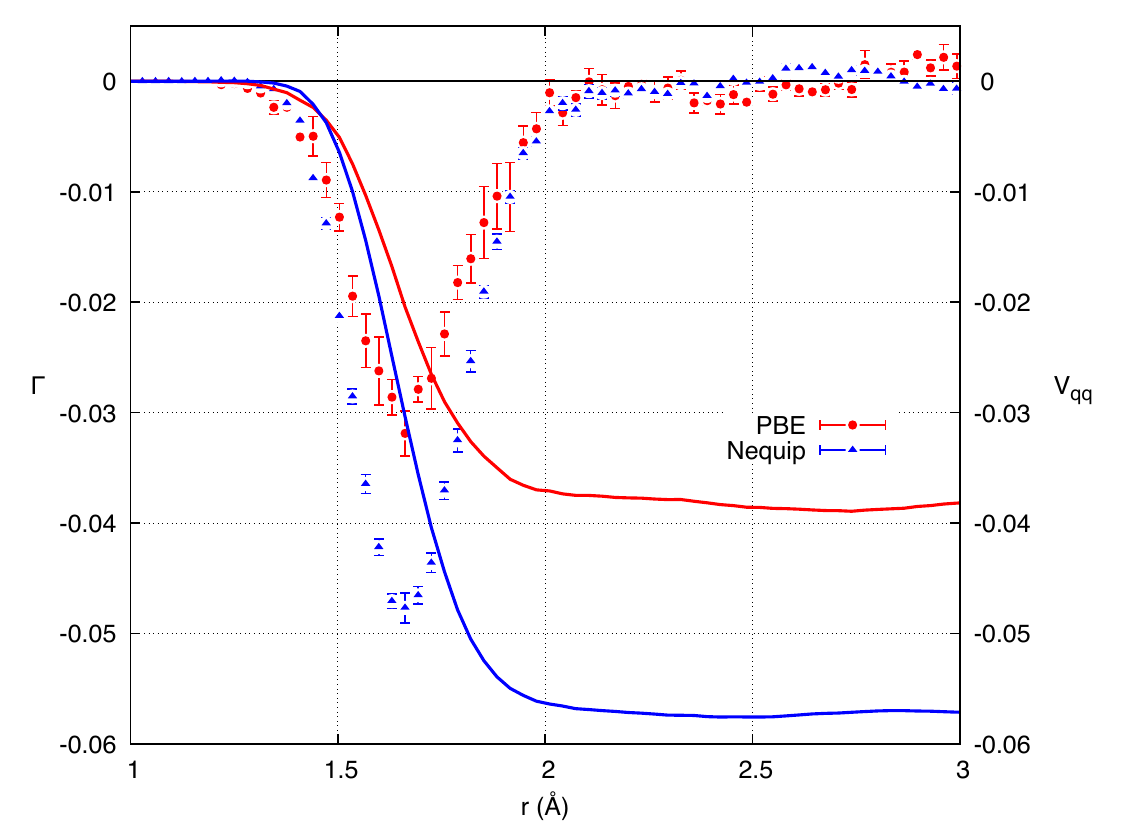}
\caption{Comparison between AIMD (red circles and line) and Nequip (blue triangles and line) correlation functions describing the quadrupole interaction as defined in Eq. \ref{eq:gamma} at 1100K, $r_s= 1.466$ with 200 atoms. The lines show the integrated quadrupole-quadrupole interaction energy $V_{qq}(r)$ assuming a quadrupole moment $\Theta=1$.} 
\label{fig:PBE_MLIP_quad_comparison}
\end{figure}

Figure~\ref{fig:PBE_MLIP_quad_comparison} shows this orientational correlation averaged over all pairs of molecules comparing data from AIMD with data from the Nequip model. 
We see that the Nequip model has 50\% larger angular correlation of adjacent molecules than the AIMD model under these conditions.
The quadrupole-quadrupole contribution to the average energy will be given by the integral of Eq. (\ref{eq:Vqq}), $V_{qq}(r)=\int_0^r dr'\mathcal{V}(r')$, represented in figure \ref{fig:PBE_MLIP_quad_comparison} by the continuous curves (with a molecular quadrupole $\Theta=1$). The Nequip interaction is 1.46 times larger than the PBE estimate.

\section{Molecular Cluster Analysis by Persistence Time}
\label{sec:persistence}

To robustly determine the formation of H$_2$ molecules, we extend the distance-based cluster analysis of previous section~\cite{pierleoniLocalStructureDense2018} to include temporal information from a MD trajectory.
We assume that atoms participating in a well-defined H$_2$ molecule should complete at least a few cycles of molecular vibration before dissociating.
Therefore, we can define a molecule by the persistent oscillation of the distance between two atoms located close to each other.

In a practical algorithm, we first estimate the probability, $p_B(r)$, that a pair of atoms is bonded knowing only their instantaneous separation. This function is determined by fitting the ratio of the molecular and atomic pair correlation functions approximated as:
\begin{equation}
\label{eq:bond_weight}
p_B(r) = \frac{1+\tanh(C(r_0-r))}{2},
\end{equation}
where  $C=4.29$ ~\AA$^{-1}$ and $r_0=0.766$~\AA.  
$p_B(r)$ vanishes when the separation is large and rises rapidly to one as $r$ decreases to less than the molecular bond length.
For each pair of atoms $(i,j)$  we perform a moving average of this probability over a time, T,
to define the persistant bonding probability:
\begin{equation}
    B_{ij}(t)=\frac{1}{T}\int_{t-T/2}^{t+T/2}dt' p_B(r_{ij}(t')).
\end{equation}
Setting T=20fs is sufficient to capture three to four molecular vibrations using snapshots separated by $\approx$ 1 fs to resolve oscillations in the bond lengths.
When there is persistent bonding, the moving average stays around $0.5$, otherwise it is much smaller.
We define the pair (i,j) as bonded at time $t$ when $B_{ij}(t) > 0.33$. It is possible that an atom could be bonded to more than one other atom, thus defining H$_3$ etc. Such clusters were rarely observed under the conditions in this study.
Once the clusters are defined, one can label the participating atoms and observe their static and dynamical distributions as was done in fig. \ref{fig:bond-weight-snapshots}.

\section{Crystallization}
\label{sec:Crystallization}
\begin{figure}
  \includegraphics[width=\linewidth]{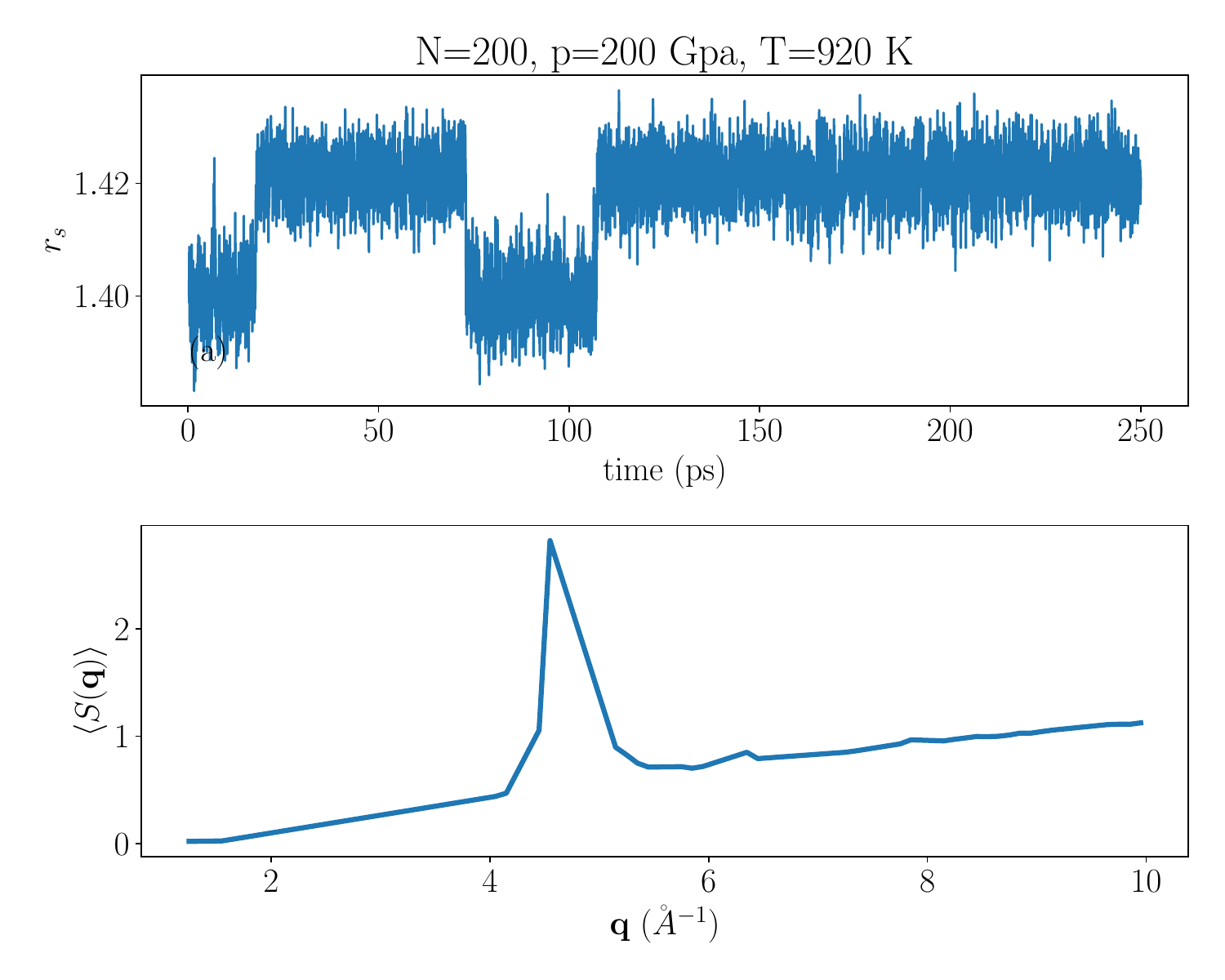}
  \caption{Instantaneous density (upper panel) and average structure factor (lower panel) for $N=200$, $P = 200$ GPa, $T=920$ K. 
  }
  \label{fig:structure_factor}
\end{figure}
This article shows the existence of the LLPT for temperatures in the 1050-1300 K range and pressures from 150-200 GPa.
But hydrogen modeled by the PBE functional at these pressures and temperatures is close to the melting line\cite{morales2010evidence} as shown in fig.\ref{fig:phase_diagram_melting}. We have to ensure that our evidence for phase transition comes from the LLPT and not from crystallization.

Figure~\ref{fig:density_traj_2048_175_1050}(c) shows the absence of Bragg peaks in the structure factor at T=1050 K,
\begin{equation}
    S(\textbf{q}) = \frac{1}{N} \left\langle \sum_k^N \sum_l^N \exp(i \textbf{q} \cdot (\textbf{r}_k - \textbf{r}_l ) )\right\rangle,
\end{equation}
 computed with the python package Freud\cite{freud2020}.

In fig.~\ref{fig:structure_factor} we show that even at the temperature of T=920 K, far below the melting temperature and with larger fluctuations due to a smaller number of particles (N=200), the structure factor has no Bragg peaks.
 Because of the long-range crystalline order, hysteresis in the liquid-solid transition is more pronounced than for the LLPT. 
 In comparison with the melting/freezing, the switches between molecular and atomic liquids are more frequent.

\begin{figure}
\includegraphics[width=\linewidth]{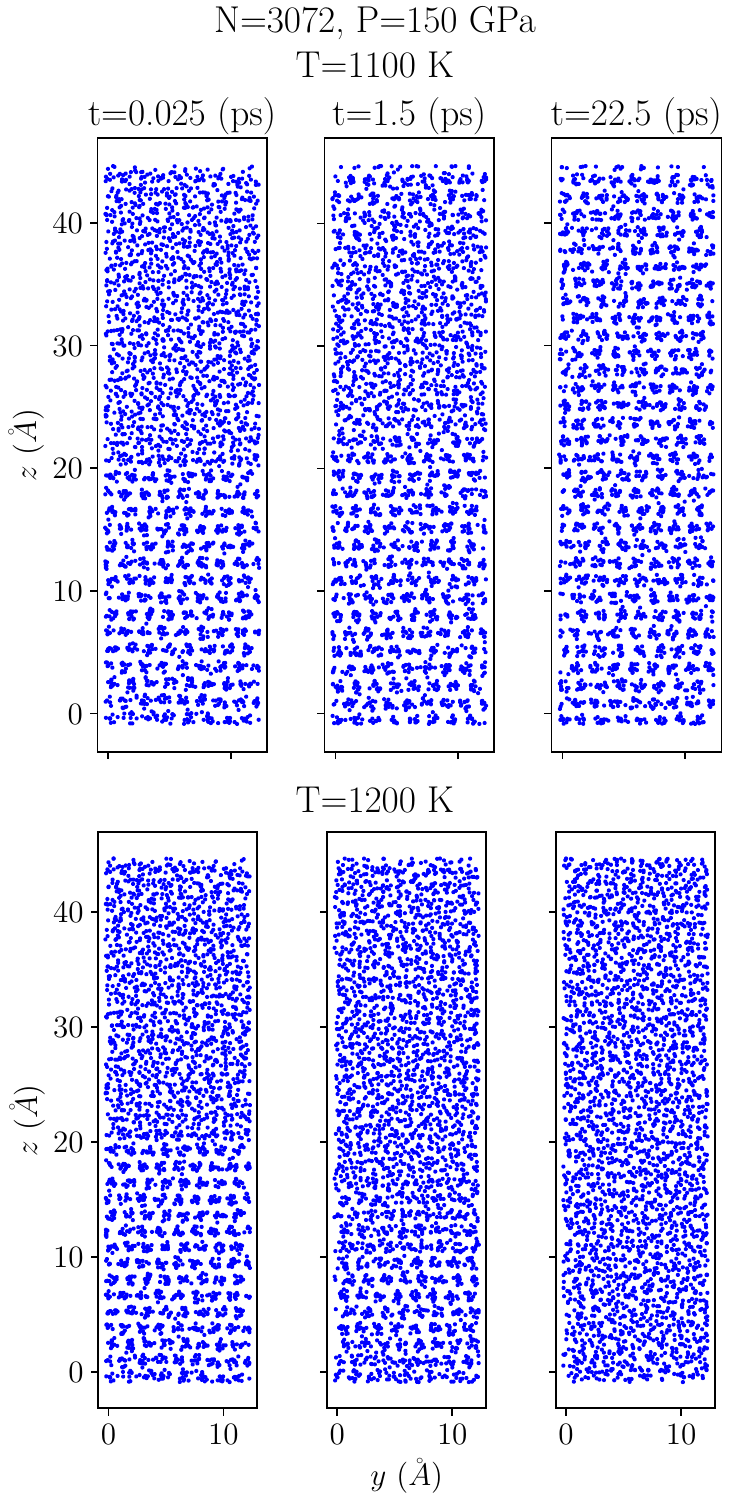}
\caption{Time evolution of a two-phase system at 150 GPa and 1100 K (top) and 1200 K (bottom) for 3072 atoms. Snapshots of the atomic position are projected onto the y-z plane to highlight the crystalline order. The left snapshots are the initial configurations, liquid for z$<20~\AA$ and solid below. The middle column shows the density after 1.5 ps. The last column after 22.5 ps when both system have equilibrated to a single phase at a later time: the top a crystal, the bottom a liquid.}
\label{fig:two_phase}
\end{figure}

Figure~\ref{fig:two_phase} shows the time-evolution two simulations initialized in a two-phase state, as in ref.~\onlinecite{goswami2024high}.
The system consist of 3072 atoms in an orthorhombic box of $\sim40~\AA$ in the $z$ direction, where the lower half of the system is initialized in a solid hcp state and the upper half in a molecular liquid state.

We then perform machine learning MD in a NPT-ensemble with periodic boundary conditions.
Figure~\ref{fig:two_phase} shows snapshots of the positions projected in the $(y, z)$ plane for 1100 K (top) and 1200 K (bottom) at 150 GPa.
On the left column, the system is shown right after initialization and the two simulations are indistinguishable at this stage.
At intermediate times, in the middle column, the number of hcp layers increases at 1100 K and decreases at 1200 K.
At later times, the two liquid-solid interfaces have disappeared and the system is in a single homogeneous phase at both 1100 K and 1200 K, the former being a hcp solid and the latter a molecular liquid giving
upper and lower bounds for the melting temperature.
Such simulations were repeated with temperatures in between 1100 and 1200K to get more precise bounds.
We then repeated the procedure every 25 GPa between 100 and 200 GPa and plotted the bounds in fig.\ref{fig:phase_diagram_melting}.

\end{appendix}

\end{document}